\documentclass[aps,showpacs,onecolumn,superscriptaddress,showkeys]{revtex4}
\usepackage{graphics,color}
\usepackage{epsfig}
\usepackage{dcolumn}
\usepackage{bm}
\usepackage{epsfig}
\usepackage{mathrsfs}
\usepackage{amsfonts}
\usepackage{graphicx}
\usepackage{amsmath}
\usepackage{amssymb}
\usepackage{subfigure}

\begin{document}

\title{A study on the relations between the topological parameter and entanglement}

\author{Chunfang Sun }
 \email[email:] {suncf997@nenu.edu.cn}
\affiliation{School of Physics, Northeast Normal University,
Changchun 130024, People's Republic of China}
\author{Kang Xue}
  \email[email:] {Xuekang@nenu.edu.cn}
\affiliation{School of Physics, Northeast Normal University,
Changchun 130024, People's Republic of China}
\author{Gangcheng Wang}
\affiliation{School of Physics, Northeast Normal University,
Changchun 130024, People's Republic of China}
\author{Chunfeng Wu}
\affiliation{Department of Physics, National University of
Singapore, 2 Science Drive 3, Singapore 117542} \affiliation{Centre
for Quantum Technologies, National University of Singapore, 3
Science Drive 2, Singapore 117543}

\begin{abstract}

In this paper, some relations between the topological parameter $d$
and concurrences of the projective entangled states have been
presented. It is shown that for the case with $d=n$, all the
projective entangled states of two $n$-dimensional quantum systems
are the maximally entangled states (i.e. $C=1$). And for another
case with $d\neq n$, $C$ both approach $0$ when $d\rightarrow
+\infty$ for $n=2$ and $3$. Then we study the thermal entanglement
and the entanglement sudden death (ESD) for a kind of Yang-Baxter
Hamiltonian. It is found that the parameter $d$ not only influences
the critical temperature $T_{c}$, but also can influence the maximum
entanglement value at which the system can arrive at. And we also
find that the parameter $d$ has a great influence on the ESD.
\end{abstract}

\pacs{02.10.Kn, 03.65.Ud, 03.67.Mn}

\keywords{Topological parameter;  Entanglement; Temperley-Lieb
algebra; Yang--Baxter system }

 \maketitle

\section{Introduction}
Quantum Entanglement(QE)\cite{R.F. Werner}, the most surprising
nonclassical property of quantum system, provides a fundamental
resource in realizing quantum information and quantum computers
\cite{M. Nielsen} and is widely exploited in quantum
cryptography\cite{A.K. Ekert}, dense coding, teleportation
\cite{C.H. Bennett}. It has been clarified that the entanglement of
a quantum state is one of the most important properties not only in
quantum information science but also in condensed matter physics.
The thermal entanglement has been investigated in the system of the
Heisenberg XXX \cite{XXX,XXX1}, XX \cite{XX}, XXZ \cite{XXZ}, and
the Ising \cite{D.G} models. Recently in Ref.\cite{T. Yu1} it has
been shown that there exists a certain class of two-qubit states
which display a finite entanglement decay time. This phenomenon is
aptly called ESD and cannot be predicted from quantum decoherence
which is an asymptotic phenomenon. It has received a lot of
attentions both theoretically and experimentally\cite{T. Yu1,T.
Yu2,Z.F,S.G,M. P}.

The Temperley-Lieb algebra(TLA) first appeared in statistical
mechanics as a tool to analyze various interrelated lattice
models\cite{TLA} and was related to link and knot
invariants\cite{wda}. Either algebraically by generators and
relations as in Jones¡¯ original presentation\cite{Jones}, or as a
diagram algebra modulo planar isotopy as in Kauffman¡¯s
presentation\cite{Kauffman3}, the TLA has always hitherto been
presented as a quotient of some sort. Recently in Ref.\cite{zhang},
the TLA is found to present a suitable mathematical framework for
describing quantum teleportation, entangle swapping, universal
quantum computation and quantum computation flow. In a very recent
work\cite{Abramsky3}, Abramsky traced some of the surprising and
beautiful connections from knot theory to logic and computation via
quantum mechanics. However, the physical meaning of the important
topological parameter $d$ (describing the unknotted loop ¡°$
\bigcirc $¡± in topology) is still unclear. Motivated by this, in
this paper we focus on studying the relations between the parameter
$d$ and entanglement to explore what role do the parameter $d$ play
in the entanglement.

The paper is organized as follows: In Sec.2, we study the relations
between the topological parameter $d$ and concurrences of the
projective entangled states. It is shown that for the case with
$d=n$, all the projective entangled states of two $n$-dimensional
quantum systems are the maximally entangled states (i.e. $C=1$). And
for another case with $d\neq n$, $C$ both approach $0$ when
$d\rightarrow\infty$ for $n=2$ and $3$.  In Sec.3, the thermal
entanglement for a kind of Yang-Baxter Hamiltonian related to the
TLA is investigated. We find that the parameter $d$ has great
influences on the thermal entanglement. It not only influences the
critical temperature $T_{c}$, but also can influence the maximum
entanglement value at which the system can arrive at. In Sec.4, the
ESD for the same Yang-Baxter Hamiltonian is investigated. It is
found that the parameter $d$ has a great influence on the ESD. A
summary is given in the last section.

\section{ some relations between the parameter $d$ and $C$}\label{sec2}

In this section, we first obtain the projective entangled state
$|\Psi_{i,i+1}\rangle$ of two $n$-dimensional quantum systems, which
contains the topological parameter $d$. Then we come to investigate
the concurrences $C$ of the states $|\Psi_{i,i+1}\rangle$ to explore
some relations between the parameter $d$ and $C$.

In order to keep the paper self-contained, we first briefly review
the theory of the TLA\cite{TLA}. It is a unital algebra generated by
$U_{i}$( $i=1,2,...,N-1$) which satisfy the following relations,
\begin{eqnarray}\label{1}
 U_{i}^{2}&=&dU_{i}  \nonumber\\
 U_{i}U_{j}U_{i}&=&U_{i}~~~~~~~~~~|i-j|=1  \nonumber\\
U_{i}U_{j}&=&U_{j}U_{i} ~~~~~~~|i-j|>1
\end{eqnarray}
where $d$ ($ d\in\mathbb{C}$ and $ d\neq 0$) is the unknotted loop
¡°$ \bigcirc $¡± in the knot theory which does not depend on the
sites of the lattices. The notation $U_{i}\equiv U_{i,i+1}$ is used,
$U_{i,i+1}$ is short for $1_{1}\otimes \cdots \otimes 1_{i-1}\otimes
U_{i,i+1}\otimes 1_{i+2} \otimes \cdots \otimes1_{N}$, and $1_{j}$
represents the unit matrix of the $j$-th particle. These relations
are diagrammatically represented in Fig. 1.
\begin{figure}
\includegraphics[scale=0.4]{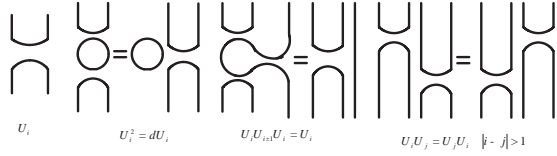}
\caption{Diagrammatic the TLA}\label{fig1}
\end{figure}

 In the following, we write the $n^{2}\times n^{2}$
$(n=2,3,...,n)$ matrix $U$ as a form of projectors in the tensor
product of two nearest $n$-dimensional quantum spaces as follows,
\begin{equation}\label{U}
U_{i,i+1}=d|\Psi_{i,i+1}\rangle\langle\Psi_{i,i+1}|,
\end{equation}
where $|d|^{\frac{1}{2}}|\Psi_{i,i+1}\rangle$ describes $\bigcup$
and $|d|^{\frac{1}{2}}\langle\Psi_{i,i+1}|$ describes $\bigcap$ in
topology. Although the parameter $d$ can be arbitrary, in this paper
we restrict ourselves on $d> 0$ for convenience. The projective
entangled state $|\Psi_{i,i+1}\rangle$ of two $n$-dimensional
quantum systems, which contains the topological parameter $d$, takes
of the following form,
\begin{eqnarray}\label{state}
\begin{array}{ll}
|\Psi_{i,i+1}\rangle=\sum_{\lambda,\mu=0}^{n-1}\alpha_{\lambda\mu}|\lambda\rangle_{i}|\mu\rangle_{i+1}.
\end{array}
\end{eqnarray}
where $|\lambda\rangle_{i}$ and $|\mu\rangle_{i+1}$ are the
orthonormal bases of the Hilbert spaces $i$ and $i+1$ respectively,
and $\alpha_{\lambda\mu}$'s are complex numbers satisfying the
normalization condition
$\sum_{\lambda,\mu=0}^{n-1}|\alpha_{\lambda\mu}|^{2}=1$. And we set
in each row $\lambda$ and each column $\mu$ of the matrix $\alpha$
there is a single nonzero element. The generators can be written as,
\begin{equation}\label{5}
(U_{i,i+1})^{\lambda\mu}_{\lambda^{'}\mu^{'}}=d
\alpha_{\lambda\mu}\alpha_{\lambda^{'}\mu^{'}}^{\ast}~~~~~~
\lambda,\mu,\lambda^{'},\mu^{'}=0,1,2,...,n-1.
\end{equation}
By calculation, it is easy to see that the first relation of
Eq.(\ref{1}) is automatically satisfied. In order to satisfy the
second relation of Eq.(\ref{1}), the fulfilled conditions read,
\begin{equation}\label{6}
\left\{
\begin{array}{ll}
d^{2}\sum_{\lambda,\nu,\sigma=0}^{n-1}\alpha_{\nu\lambda}^{\ast}\alpha_{\lambda\mu}\alpha_{\nu\sigma}\alpha_{\sigma\beta}^{\ast}=\delta_{\mu\beta},\\
&\\
 d^{2}\sum_{\lambda,\nu,\sigma=0}^{n-1}\alpha_{\mu\lambda}\alpha_{\lambda\nu}^{\ast}\alpha_{\beta\sigma}^{\ast}\alpha_{\sigma\nu}=\delta_{\mu\beta},
\end{array}
\right.
\end{equation}
where $\mu=0,1,2,...,n-1$. By this limited conditions Eq.(\ref{6}),
the projective entangled state $|\Psi_{i,i+1}\rangle$ of two
$n$-dimensional quantum systems and the corresponding topological
parameter $d$ can be determined. Via Eq.(\ref{U}), the corresponding
$n^{2}\times n^{2}$ matrix $U$ can also be obtained. Next via two
classes of the parameter $d$, we come to study  the relations
between the parameter $d$ and the concurrences $C$ of the
corresponding states $|\Psi_{i,i+1}\rangle$.

\subsection{Example I: the case with the parameter $d=n$}

In example $I$, we will discuss a series of the generalized
$n^{2}\times n^{2}$ ($n=2,3,...,n$) matrix $U$ with the topological
parameter $d=n$.

For the case with $\lambda=\mu$ and
$\lambda^{'}=\mu^{'}$($\lambda,\mu,\lambda^{'},\mu^{'}=0,1,2,...,n-1$)
in the tensor product of two nearest $n$-dimensional quantum spaces,
via Eq.(\ref{6}) and Eq.(\ref{state}), the corresponding state is
\begin{equation}\label{n}
|\Psi\rangle=\sum_{\lambda=0}^{n-1}\frac{1}{\sqrt{n}}e^{ik_{\lambda\lambda}}|\lambda\lambda\rangle,
\end{equation}
where the topological parameter $d=n$ and the parameters
$k_{\lambda\lambda}$ are arbitrary real. By means of concurrence, we
study these entangled states. In Ref.\cite{Albeverio}, the
generalized concurrence (or the degree of entanglement\cite{Hill})
for two qudits is given by,
\begin{equation}\label{C}
   C=\sqrt{\frac{n}{n-1}(1-I_{1})},
\end{equation}
where
$I_{1}=Tr[\rho_{A}^{2}]=Tr[\rho_{B}^{2}]=|\kappa_{0}|^{4}+|\kappa_{1}|^{4}+\cdots+|\kappa_{n-1}|^{4}$,
with $\rho_{A}$ and $\rho_{B}$ are the reduced density matrices for
the subsystems, and $\kappa_{j}$'s($j=0,1,\ldots,n-1$) are the
Schmidt coefficients. Then we can obtain the generalized concurrence
of the state $|\Psi\rangle$ (\ref{n})  as follows,
\begin{equation} \label{20}
C=1.
\end{equation}
It is interesting that for the series of $n^{2}\times n^{2}$ matrix
$U$ with the topological parameter $d=n$, all the projective states
$|\Psi\rangle$ have the maximum entanglement. The state
$|\Psi\rangle$ in Eq. (\ref{n}) can be considered as a
straightforward generalization of the symmetric Bell state
$|\Phi^{+}\rangle=\frac{1}{\sqrt{2}}(|00\rangle+|11\rangle)$ when
$e^{ik_{\lambda\lambda}}=1$.

\subsection{Example II: the case with the parameter $d\neq n$ }
 In example $II$, we will discuss another class of  the $n^{2}\times n^{2}$ matrix $U$ with
the topological parameter $d\neq n$. Because in this case we can't
obtain the general generalized $n^{2}\times n^{2}$ matrix $U$, we
will study the cases with $n=2$ and $n=3$.

For the case with $n=2$, via Eq.(\ref{6}) and Eq.(\ref{state}), the
corresponding state is
\begin{equation}\label{state 1}
\begin{array}{ll}
|\Psi\rangle=\frac{1}{\sqrt{1+q^{2}}}(q
e^{ik_{01}}|01\rangle+e^{ik_{10}}|10\rangle),
\end{array}
\end{equation}
where the topological parameter $d=q+q^{-1}$ and
$k_{01},k_{10},q\in$ real. Hereafter, $q> 0$. The generalized
concurrence of the state (\ref{state 1}) is,
\begin{equation} \label{concurrence2}
C=\frac{2}{d}, ~~~~~~~~where ~~~d\geq 2.
\end{equation}
 For the case with $n=3$, via Eq.(\ref{6}) and
Eq.(\ref{state}), there are three sets of solutions and the
corresponding states are,
\begin{equation}\label{state 2}
\begin{array}{ll}
|\Psi\rangle^{(1)}=\frac{1
}{\sqrt{1+q+q^{2}}}(qe^{ik_{02}}|02\rangle+\sqrt{q}e^{ik_{11}}|11\rangle+e^{ik_{20}}|20\rangle),\\
&\\
|\Psi\rangle^{(2)}=\frac{1
}{\sqrt{1+q+q^{2}}}(qe^{ik_{01}}|01\rangle+e^{ik_{10}}|10\rangle+\sqrt{q}e^{ik_{22}}|22\rangle),\\
&\\
|\Psi\rangle^{(3)}=\frac{1
}{\sqrt{1+q+q^{2}}}(\sqrt{q}e^{ik_{00}}|00\rangle+qe^{ik_{12}}|12\rangle+e^{ik_{21}}|21\rangle),
\end{array}
\end{equation}
where the parameter $d=q+q^{-1}+1$ and $k_{\lambda\mu}$ $\in$ real
($\lambda,\mu=0,1,2$). All their concurrences are the same as,
\begin{equation} \label{concurrence3}
C=\sqrt{\frac{3}{d}},~~~~~~~~where~~~ d\geq 3.
\end{equation}

Via these two examples, it is shown that there are
 some relations between the topological parameter
 $d$ and concurrences $C$ of the entangled states. In other words, the parameter $d$ has great influences on the
 entanglement. Example $I$ and Example $II$(i.e., $q=1$) show that for the series of the parameter $d=n$, all the projective states $|\Psi\rangle$ of two $n$-dimensional quantum systems
are the maximally entangled states (i.e., $C=1$). For another class
of the parameter $d\neq n$ (i.e., $q\neq1$) in example $II$, via
investigating the cases with $n=2$ and $n=3$, it is found that the
concurrences $C$ both decrease when $d$ goes up, and it approaches
$0$ when $d\rightarrow +\infty$, as shown in Fig. 2. We guess that
for the generalized $n^{2}\times n^{2}$ matrix $U$ with the
parameter $d\neq n$, the conclusion, which is that when the
parameter $d\rightarrow +\infty$, $C$ approaches $0$, is also
correct. Another fact in Fig. 2 is that for the same value of loop
$d$, the concurrence of the entangled two-qutrit states (\ref{state
2}) is always larger than the concurrence of the entangled two-qubit
states (\ref{state 1}). This means that the concurrence not only
depends on the topological parameter $d$, but also depends on the
dimension $n$.
\begin{figure}
\includegraphics[scale=0.5]{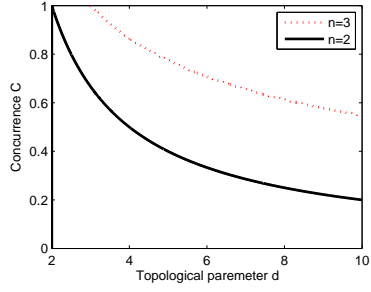}
\caption{The concurrence is plotted versus the parameter $d$. The
solid line corresponds to $C=\frac{2}{d}$ for $n=2$, and the dotted
line corresponds to $C=\sqrt{\frac{3}{d}}$ for $n=3$}\label{fig2}
\end{figure}

\section{Thermal entanglement in a YANG-BAXTER SYSTEM}\label{sec4}

 In this section, we come to study the thermal
entanglement for a kind of Yang-Baxter Hamiltonian, which is related
to the TLA for $n=2$, to explore the influences of the parameter $d$
on the thermal entanglement.

By substituting Eq.(\ref{state 1}) into Eq.(\ref{U}) for $n=2$, in
the standard basis $\{|00\rangle, |01\rangle, |10\rangle,
|11\rangle\}$, the $4\times 4$ matrix $U$ is,
\begin{eqnarray}
 U=\left(
  \begin{array}{cccc}
   0 & 0 & 0 & 0  \\
0 & q & e^{i\varphi} & 0  \\
0 & e^{-i\varphi} & q^{-1} & 0  \\
0 & 0 & 0 &  0  \\
\end{array}
\right),
\end{eqnarray}
with the topological parameter $d=q+q^{-1}$ and
$\varphi=k_{01}+k_{10} \in $ real.

 As is known,
the Yang-Baxter equation (YBE)\cite{yang,baxter,drin} is given by,
\begin{equation}
\breve{R}_{i}(x)\breve{R}_{i+1}(xy)\breve{R}_{i}(y)=\breve{R}_{i+1}(y)\breve{%
R}_{i}(xy)\breve{R}_{i+1}(x) , \label{YBE}
\end{equation}
where $x$ and $y$ are spectrum parameters. Via the trigonometric
Yang-Baxterization approach\cite{ckg2}, it gives

\begin{eqnarray}
 \breve{R}(x)=[q^{2}+q^{-2}-(x^{2}+x^{-2})]^{-\frac{1}{2}}[(q
 x-q^{-1}x^{-1})I
 -(x-x^{-1})U], \nonumber\\
\breve{R}^{-1}(x)=[q^{2}+q^{-2}-(x^{2}+x^{-2})]^{-\frac{1}{2}}[(qx^{-1}-q^{-1}x)I
 +(x-x^{-1})U].
\end{eqnarray}
It is easy to check that
$\breve{R}^{+}(x)=\breve{R}^{-1}(x)=\breve{R}(-x)$ for $
x=e^{i\theta}$, where $\theta\in$ real. It is worth to mention that
in this paper, the real parameters $\theta$ and $\varphi$ are
time-independent.

  Here we study a original Hamiltonian describing two spin-1/2
  particles (particle 1 and 2) interaction,
$$H_{0}=\mu_{1}S_{1}^{z}+\mu_{2}S_{2}^{z}+gS_{1}^{z}S_{2}^{z},$$
where $\mu_{i}$ ($i=1,2$) represent external magnetic field and $g$
is the interaction of $z$-component of two-qubit spins. Taking the
Schr$\ddot{o}$dinger equation $i\hbar
\partial|\Psi\rangle/\partial
t=H|\Psi\rangle$ into account, where
$|\Psi\rangle=\check{R}(x)|\Psi_{0}\rangle$ and $|\Psi_{0}\rangle$
is the eigenstate of $H_{0}$, one can get a new Hamiltonian  as
$H(\theta,\varphi)=\check{R}(x)H_{0}\breve{R}^{-1}(x)$\cite{sun},
where the real parameters $\theta$ and $\varphi$ are
time-independent. For convenience, we let $x=i$ (i.e.,
$\theta=\frac{\pi}{2}$). Then we arrive at a new Hamiltonian,
\begin{eqnarray}\label{H}
H=(B+J(1-\frac{8}{d^{2}}))S^{z}_{1}+(B-J(1-\frac{8}{d^{2}}))S^{z}_{2}
+gS^{z}_{1}S^{z}_{2}
-\frac{4J\sqrt{d^{2}-4}}{d^{2}}(e^{i\varphi}S^{+}_{1}S^{-}_{2}+e^{-i\varphi}S^{-}_{1}S^{+}_{2})
,
\end{eqnarray}
where $B=\frac{\mu_{1}+\mu_{2}}{2}$ and
$J=\frac{\mu_{1}-\mu_{2}}{2}$, and $S^{\pm}_{i}=S^{x}_{i}\pm i
S^{y}_{i}$ are raising and lowering operators respectively for the
$i$-th particle. Specifically, we find that when $\varphi=\pi$, this
model is the two-qubit anisotropic Heisenberg $XXZ$ model under an
inhomogeneous magnetic field. $B \geq 0$ is restricted, and the
magnetic fields on the two spins have been so parameterized that
$J(1-8/d^{2})$ controls the degree of inhomogeneity. For the system
(\ref{H}), its corresponding eigenstates read
 $|\Psi_{1}\rangle=|00\rangle$, $|\Psi_{2}\rangle=|11\rangle|$,
 $|\Psi_{3}\rangle=\frac{2}{d}(\frac{-\sqrt{d^{2}-4}e^{i\varphi}}{2}|01\rangle+|10\rangle)$, $|\Psi_{4}\rangle=\frac{2}{d}(|01\rangle+\frac{\sqrt{d^{2}-4}e^{-i\varphi}}{2}|10\rangle)$,
with corresponding energies
 $E_{1}=B+\frac{g}{4}$, $E_{2}=-B+\frac{g}{4}$,
$E_{3}=J-\frac{g}{4}$, $E_{4}=-J-\frac{g}{4}$.

Next to quantify the entanglement of formation of a mixed state
$\rho$ of two qubits, we use the Wootters concurrence \cite{W.K}
defined as,
\begin{eqnarray}\label{c}
C(t)
=\max\{0,\sqrt{\lambda_{1}}-\sqrt{\lambda_{2}}-\sqrt{\lambda_{3}}-\sqrt{\lambda_{4}}\},
\end{eqnarray}
where  $\{\lambda_{i}\}$ are the eigenvalues of the matrix
$\rho(\sigma_{y}^{A}\otimes
\sigma_{y}^{B})\rho^{\ast}(\sigma_{y}^{A}\otimes \sigma_{y}^{B})$,
with $\rho^{\ast}$ denoting complex conjugation of the matrix $\rho$
and $\sigma_{y}^{A/B}$ are the Pauli matrices for atoms A and B.
When spin chains are subjected to environmental disturbance, they
inevitably become thermal equilibrium states. The thermal state at
finite temperature $T$ is $\rho(T)=\frac{1}{Z}\exp(-\frac{H}{kT})$,
where $Z={\rm Tr}[\exp(-\frac{H}{kT})]$ is the partition function
and $k$ is the Boltzmann constant. For simplicity, we write $k=1$.
By calculation, the density matrix $\rho(T)$ of the system (\ref{H})
can be written as,
\begin{eqnarray}
\rho(T)=\frac{1}{2(\cosh\frac{B}{T}+e^{\frac{g}{2T}}\cosh\frac{J}{T})}
\left(
  \begin{array}{cccc}
    e^{\frac{-B}{T}} & 0  & 0 & 0 \\
    0 & e^{\frac{g}{2T}}(\cosh\frac{J}{T}-(1-\frac{8}{d^{2}})\sinh\frac{J}{T}) & \frac{4\sqrt{d^{2}-4}}{d^{2}}e^{\frac{g}{2T}}\sinh\frac{J}{T}e^{i\varphi} & 0 \\
   0  & \frac{4\sqrt{d^{2}-4}}{d^{2}}e^{\frac{g}{2T}}\sinh\frac{J}{T}e^{-i\varphi} & e^{\frac{g}{2T}}(\cosh\frac{J}{T}+(1-\frac{8}{d^{2}})\sinh\frac{J}{T}) & 0 \\
    0 & 0 & 0 & e^{\frac{B}{T}} \\
  \end{array}
\right).
\end{eqnarray}
The concurrence is calculated as,
\begin{eqnarray}\label{c2}
C = \max\left(
\frac{\frac{4\sqrt{d^{2}-4}}{d^{2}}e^{\frac{g}{2T}}\sinh\frac{|J|}{T}-1}{\cosh\frac{B}{T}+
e^{\frac{g}{2T}}\cosh\frac{J}{T}},0\right).
\end{eqnarray}
Now we do the limit $T \rightarrow 0$ on the concurrence (\ref{c2}),
we obtain,
\begin{eqnarray}\label{maximum}
\lim_{T\rightarrow 0}C_1 &=& \frac{4\sqrt{d^{2}-4}}{d^{2}}~~~~~ for  ~~~~~|B|>|J|+\frac{g}{2},  \nonumber \\
&=&\frac{2\sqrt{d^{2}-4}}{d^{2}} ~~~~~ for  ~~~~~|B|=|J|+\frac{g}{2},  \nonumber \\
&=&0 ~~~~~ ~~~~~for ~~~~~~~~~~|B|<|J|+\frac{g}{2}.
\end{eqnarray}
It is worth to mention that the influences of the parameters $g$ and
$B$ on the thermal entanglement have been discussed in our
paper\cite{sun}, whose model corresponds to the topological
parameter $d=2$ (i.e., $q=1$). Here we emphasize on exploring the
parameter $d$'s influences on the thermal entanglement. From Eq.
(\ref{maximum}) we can see that at $T=0$, the entanglement vanishes
as $|B|$ crosses the critical value $|J|+\frac{g}{2}$, which means
that the critical magnetic field $B_{c}$ is independent on the
parameter $d$.
 An important point revealed by Eq. (\ref{maximum}) is that the maximum entanglement value at
which the system can arrive at, which is
$C_{max}=\frac{4\sqrt{d^{2}-4}}{d^{2}}$ for $|B|>|J|+\frac{g}{2}$ at
$T=0$, is dependent on the parameter $d$. Fig. 3 shows that when the
parameter $d=2\sqrt{2}$, the ground states
$|\Psi_{3}\rangle=\frac{1}{\sqrt{2}}(-|01\rangle+|10\rangle)$ or
$|\Psi_{4}\rangle=\frac{1}{\sqrt{2}}(|01\rangle+|10\rangle)$ both
become the maximally entangled states,  so the maximum entanglement
value $C_{max}=1$. When the parameter $d \rightarrow 2$ or $d
\rightarrow +\infty$, the ground states $|01\rangle$ or $|10\rangle$
both have no entanglement, then the maximum entanglement value
$C_{max}=0$. Another important character revealed by Eq. (\ref{c2})
is that the critical temperature $T_{c}$, which is determined by the
nonlinear equation
$\frac{4\sqrt{d^{2}-4}}{d^{2}}e^{\frac{g}{2T_{c}}}\sinh\frac{|J|}{T_{c}}=1$,
is also dependent on the parameter $d$. From Fig. 4, it is shown
that when $d=2\sqrt{2}$, $T_{c}$ arrive at the maximum value (i.e.,
$T_{c}$ is about 1.5). When $d \rightarrow 2$ or $d \rightarrow
+\infty$, all the four eigenstates are unentangled states, so the
critical temperature $T_{c}=0$. Thus we can obtain a higher
entanglement at a fixed temperature via changing the values of the
parameter $d$.

\begin{figure}
\includegraphics[scale=0.4]{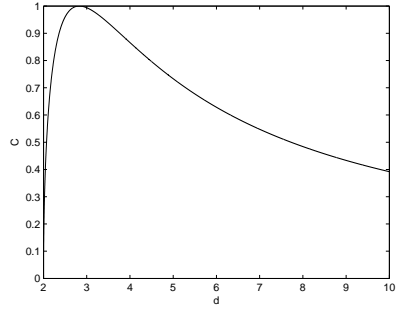}
\caption{The maximum entanglement value $C_{max}$ is plotted versus
the parameter $d$.}\label{fig3}
\end{figure}

\begin{figure}
\includegraphics[scale=0.4]{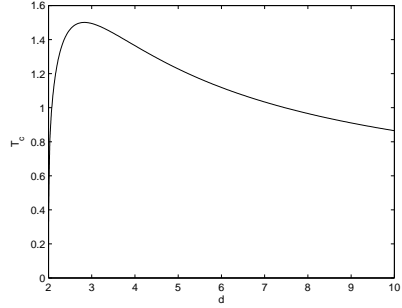}
\caption{The critical temperature $T_{c}$ is plotted versus the
parameter $d$. Coupling constant $J=1$ and the parameter
$g=1$.}\label{fig4}
\end{figure}

\section{ESD in the same YANG-BAXTER SYSTEM}

In this section, we study the ESD in the same Yang-Baxter system
(\ref{H}) to explore the influence of the parameter $d$ on the ESD.

The time evolution $U(t)=exp\{-iHt\}$ is written in the basis
$\{|00\rangle,|01 \rangle,|10\rangle, |11\rangle\}$,
\begin{eqnarray}\label{uu}
\begin{array}{lll}
  U_{11}=e^{-i(B+\frac{g}{4})t}\\
  \\
  U_{44}=e^{i(B-\frac{g}{4})t}\\
  \\
  U_{22}=e^{i\frac{g}{4}t}(\cos[J t]-i (1-\frac{8}{d^{2}})\sin[J t]),\\
   U_{33}=e^{i\frac{g}{4}t}(\cos[J t]+i (1-\frac{8}{d^{2}})\sin[J t])\\
  \\
  U_{23}=e^{i\frac{g}{4}t}\frac{4i e^{i\varphi}\sqrt{d^{2}-4}\sin[J t]}{d^{2}}, ~~~~~ U_{32}=e^{i\frac{g}{4}t}\frac{4i e^{-i\varphi}\sqrt{d^{2}-4}\sin[J
  t]}{d^{2}}.
\end{array}
\end{eqnarray}
It is convenient to choose the initial state
$\rho_{0}=\frac{1-\gamma}{4}+\gamma|\psi\rangle\langle\psi|$
$(0<\gamma\leq1)$ with $|\psi\rangle$=$sin\alpha|01\rangle
+cos\alpha|10\rangle$. It is worth to mention that in our
paper\cite{sun1}, it has been shown that in Yang-Baxter systems, the
ESD is not only sensitive to the initial condition , but also has
relations with the different Yang-Baxter systems. And it has been
found that the meaningful parameter $\varphi$ has a great influence
on the ESD. Here we emphasize on studying the influences of the
unknotted loop $d$ on the ESD. For convenience, we let the
parameters $\alpha=\frac{\pi}{4}$, $\gamma=0.5$, $J=\frac{1}{2}$ and
$\varphi=\pi$. The system model (\ref{H}) corresponds to the
two-qubit anisotropic Heisenberg $XXZ$ model under an inhomogeneous
magnetic field. Then the entanglement for
$\rho(t)=U(t)\rho_{0}U^{+}(t)$ can be given easily, and according to
Eq.(\ref{c}), the concurrence can be obtained as follows,
\begin{eqnarray}\label{c1}
C=\frac{\sqrt{(16(d^{2}-4)+(d^{2}-8)^{2}\cos
t)^{2}+d^{4}(d^{2}-8)^{2}\sin^{2} t}}{2d^{4}}-\frac{1}{4}.
\end{eqnarray}
 In Fig. 5, we give a plot of the concurrence as a function of the time $t$ and the parameter
 $d$. It is clear that in our closed Yang-Baxter system, the ESD happens in
 some special times and then the entanglement revives after a while. One can note that the topological parameter $d$ has a great
influence on the ESD when the initial condition is determinate. It
is obvious that the ESD happens only when the parameter $d$ changes
in a certain range. This means that in the Yang-Baxter system, one
can realize the ESD via changing the values of the parameter $d$
when the initial condition is determinate.

\begin{figure}
\subfigure[]{\includegraphics[scale=0.5]{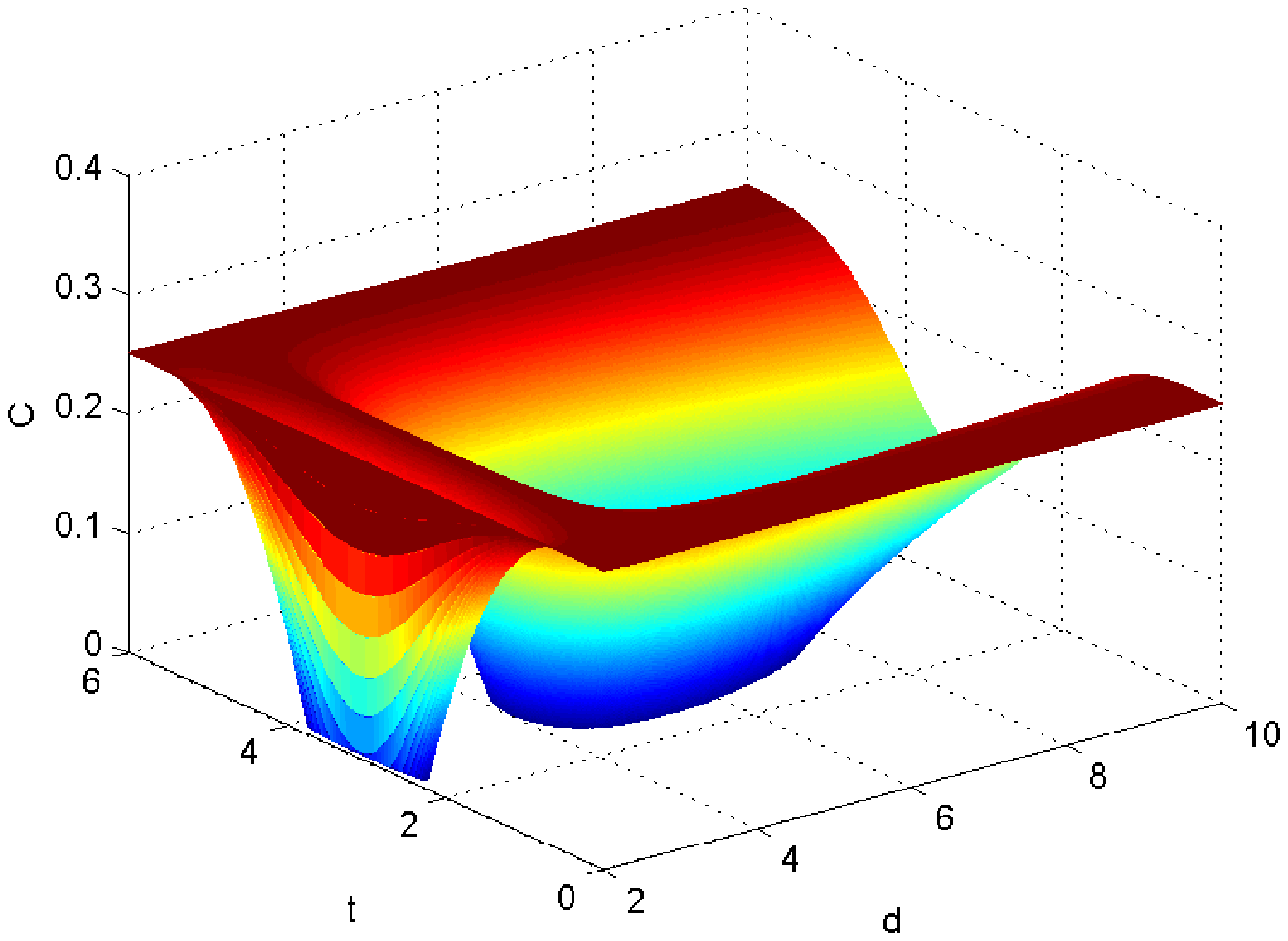}}
\subfigure[]{\includegraphics[scale=0.5]{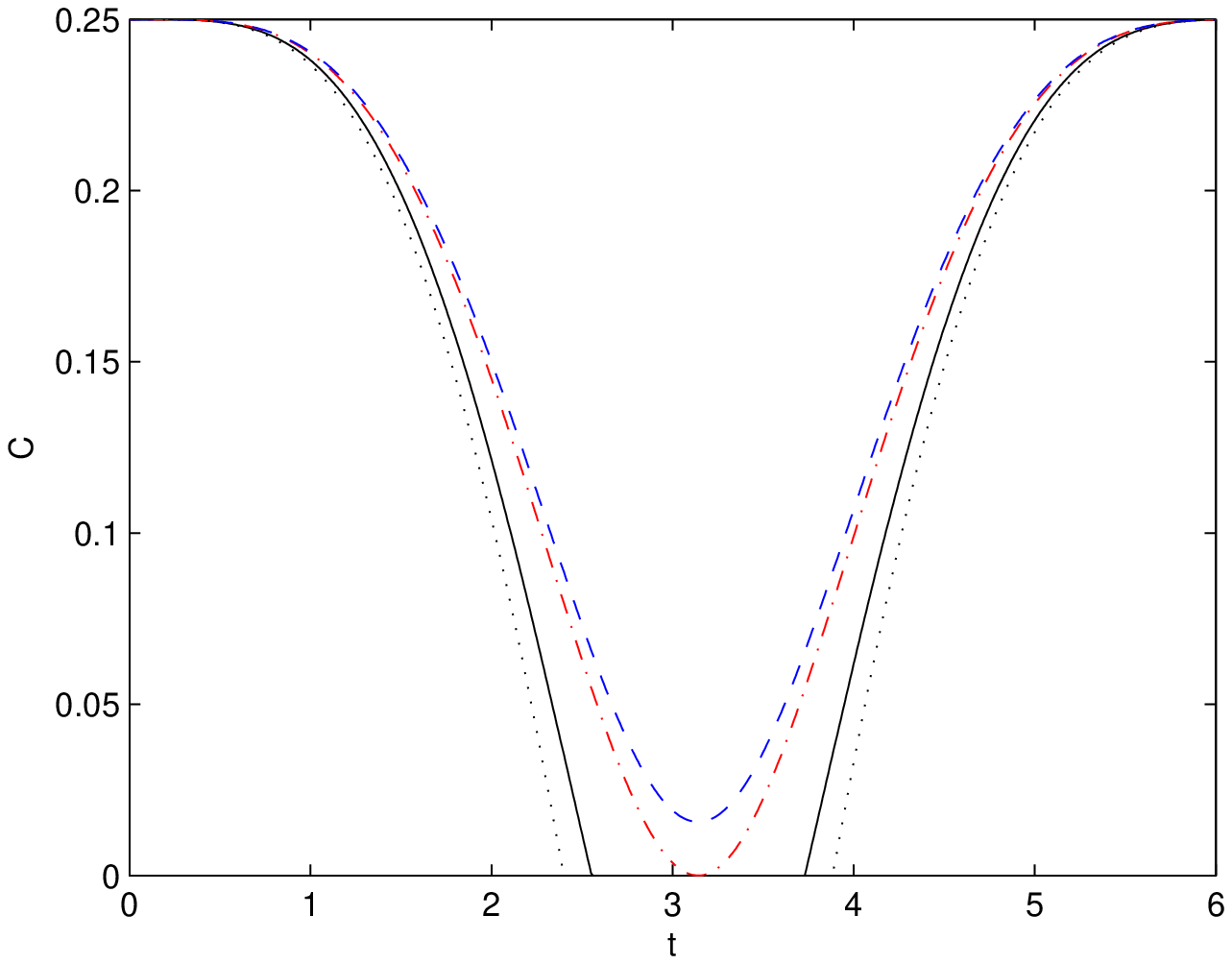}} \caption{The
concurrence $C$ is plotted versus the time $t$ and the parameter
 $d$. The figure (b),  the concurrences versus time $t$ for different
parameters $d$: $d=2.1$(solid
 line), $d=4$(dot-dashed
 line),$d=5$(dotted
line),$d=8$(dashed
 line). \label{fig5}}
\end{figure}

\section{Summary}\label{sec5}

In this paper, we have presented some relations between the
topological parameter $d$ and concurrences of the projective
entangled states. Specifically, it is shown that for the case with
the parameter $d=n$, all the projective entangled states of two
$n$-dimensional quantum systems are the maximally entangled states
(i.e. $C=1$). And for another case with the parameter $d\neq n$, via
investigating the cases with $n=2$ and $n=3$, we find $C$ both
approach $0$ when $d\rightarrow\infty$. Then we construct a kind of
Yang-Baxter Hamiltonian related to the $4\times 4$ matrix $U$, with
the topological parameter $d=q+q^{-1}$ for $n=2$. The thermal
entanglement and the ESD for the Yang-Baxter system have been
investigated. It is found that the parameter $d$ has great
influences on the thermal entanglement. It not only influences the
critical temperature $T_{c}$, but also can influence the maximum
entanglement value at which the system can arrive at. Finally we
find that the parameter $d$ also has a great influence on the ESD,
and one can realize the ESD via changing the values of the parameter
$d$ when the initial condition is determinate. It is worth to
mention that via our paper, it is obvious that the topological
parameter $d$ plays an important role in the entanglement.

\section{Acknowledgments}

We would like to thank Chengcheng Zhou for his useful discussions.
This work was supported by NSF of China (grants No. 10875026) and
NUS research (grant No. WBS: R-710-000-008-271).

\end{document}